\documentclass[a4paper, 11pt]{article}
\usepackage[a4paper,hmargin={2.5cm,2.5cm},vmargin={3cm,3cm}]{geometry}

\usepackage{amsmath,amssymb}
\usepackage{amsthm}

\usepackage{graphicx}
\usepackage{subcaption}

\newtheorem{theorem}{Theorem}
\newtheorem{lemma}{Lemma}
\newtheorem{prop}{Proposition}

\theoremstyle{remark}
\newtheorem{rem}{Remark}

\DeclareMathOperator{\sgn}{sgn}

\newcommand{\bee}{\begin{equation*}}
\newcommand{\ene}{\end{equation*}} 

\begin{document}

\title{\textbf{Anisotropic compressed sensing for non-Cartesian MRI acquisitions}}
\author{Philippe CIUCIU \footnote{Neurospin, CEA Saclay, Parietal, INRIA, 91191 Gif-sur-Yvette, France, philippe.ciuciu@cea.fr }\and
	Anna KAZEYKINA \footnote{Laboratoire de Math\'ematiques d'Orsay, Universit\'e Paris-Sud, Universit\'e
Paris-Saclay, 91405 Orsay, France,  anna.kazeykina@math.u-psud.fr} }
	\date{}

\maketitle{}

\begin{abstract}
In the present note we develop some theoretical results in the theory of anisotropic compressed sensing that allow to take structured sparsity and variable density structured sampling into account. We expect that the obtained results will be useful to derive explicit expressions for optimal sampling strategies in the non-Cartesian~(radial, spiral, etc.) setting in MRI.
\end{abstract}
	

\section{Introduction}
The mathematical problem of compressed sensing (CS) consists in recovering a sparse signal from a small number of measurements. More precisely, we wish to recover a vector $x \in \mathbb{C}^n$ from a vector of measurements $y = A x$, where $A \in \mathbb{C}^{ m \times n } $ is the sensing matrix and $ m \ll n $. The signal is said to be $s$-sparse if it has at most $s$ non-zero entries.
The recovery is usually performed by solving the following minimization problem called basis pursuit (BP):
\begin{equation}
\label{BPP}
\min_{ x \in \mathbb{C}^n, y = Ax } \| x \|_{\ell^1}.
\end{equation}

One of the classical CS results can be formulated as follows \cite{CRT, CP, FR}. Let $ A_0 \in \mathbb{C}^{ n \times n } $ satisfy the \emph{isotropy} condition: $ A_0^* A_0 = I $, and suppose that
the measurement matrix $ A $ is constructed by drawing  $ m $ random rows of $ A_0 $ in an independent uniform manner.
Define the coherence of matrix $ A $ to be $ \mu(A) = n \max_i \| a^*_i \|^2_{ \infty } $, where $ a^*_i $ are the rows of matrix $ A $. This quantity represents the coherence between the sensing and the sparsifying bases (low coherence meaning that a vector in the sparsifying basis is approximately uniformly spread in the sensing basis).
If $ m \gtrsim \mu(A) s \ln(n/\varepsilon) $, then an $s$-sparse vector $x$ can be exactly recovered by solving \eqref{BPP} with probability at least $1-\varepsilon$. 

In many applications~(including MRI), the sensing matrix $A$ is coherent, meaning that $\mu(A)$ is large. It can be shown that incoherence is typically met between the standard basis and the Fourier basis. However, natural images $x$ have sparse representations not in the pixel basis directly, but rather in wavelet bases, i.e. $ x = \Psi z $ with $z$ sparse, which are not incoherent with the Fourier basis. 

In practice, uniformly drawn measurements lead to very poor reconstructions. It was observed, however, that MR image reconstruction from undersampled frequencies could be significantly improved by drawing measurements according to variable densities strategies (VDS), preferring low to high frequencies.

VDS strategies have received a justification in the CS literature \cite{CP, CCW, CCKW}. If the measurements are drawn independently with the probability to draw the $ j $-th measure equal to $ \pi_j = \frac{ \| a^*_j \|^2_{ \infty } }{ \sum_{ j = 1 }^{ n } \| a^*_j \|^2_{ \infty } }  $, then an $ s $-sparse vector $ x $ can be reconstructed exactly from $ m $ measurements with probability at least $ 1 - \varepsilon $ provided that $m  \gtrsim\sum_{j = 1}^n \| a^*_j \|^2_{ \infty } s \ln(n/\varepsilon)$.

It was shown experimentally, however, that this result is not sufficient to explain the success of CS in applications such as MRI \cite{AHPR}. It is in particular due to the fact that in the above result we do not assume any structure~(apart from sparsity) in the signals to be recovered. A natural extension would be to consider the structured sparsity approach, where one assumes that some prior information on the support $S$ is known, e.g. sparsity by level in the wavelet domain (see \cite{AHPR} for a comprehensive theory for Fourier sampling, based on isolated measurements under a sparsity-by-levels assumption in the wavelet domain). This strategy allows to incorporate any kind of prior information on the structure of $S$ and to study its influence on the quality of CS reconstructions.

Another obstacle to applying classical CS results in a large number of practical settings is that the isolated measurements are incompatible with the physics of acquisition. For this reason, more recent relevant contributions \cite{BBW, ABB} have addressed structured VDS i.e. over sampling trajectories. This approach allows to give recovery guarantees for block-structured acquisition with an explicit dependency on the support of the vector to reconstruct and it provides many possibilities such as optimizing the drawing probability $\pi$ or identifying the classes of supports recoverable with block sampling strategies.

\begin{figure}[tb]
	\centering
	\begin{minipage}{0.27\textwidth}
		\includegraphics[width=\textwidth]{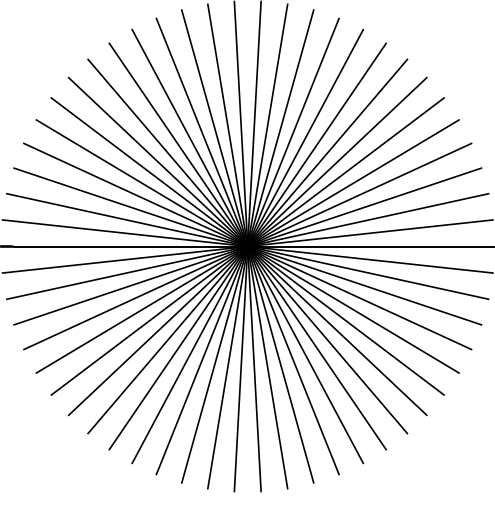}
		\subcaption{\label{subfig:non_cartesian_radial}}
	\end{minipage}
	\hspace{0.1cm}
	\begin{minipage}{0.27\textwidth}
		\includegraphics[width=\textwidth]{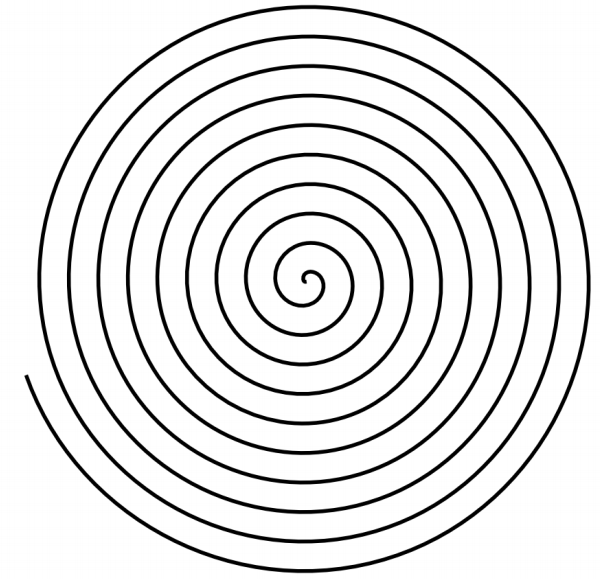}
		\subcaption{\label{subfig:non_cartesian_spiral}}
	\end{minipage}
	\hspace{0.1cm}
	\begin{minipage}{0.27\textwidth}
		\includegraphics[width=\textwidth]{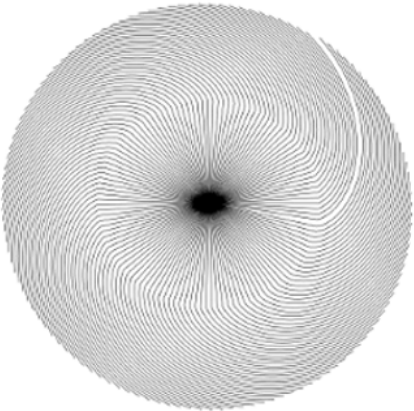}
		\subcaption{\label{subfig:non_cartesian_twirl}}
	\end{minipage}
	\\
	\vspace{0.5cm}
	\begin{minipage}{0.27\textwidth}
		\includegraphics[width=\textwidth]{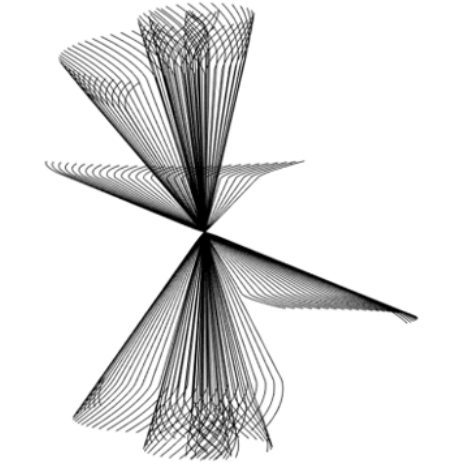}
		\subcaption{\label{subfig:non_cartesian_twist}}
	\end{minipage}
	\hspace{0.1cm}
	\begin{minipage}{0.27\textwidth}
		\includegraphics[width=\textwidth]{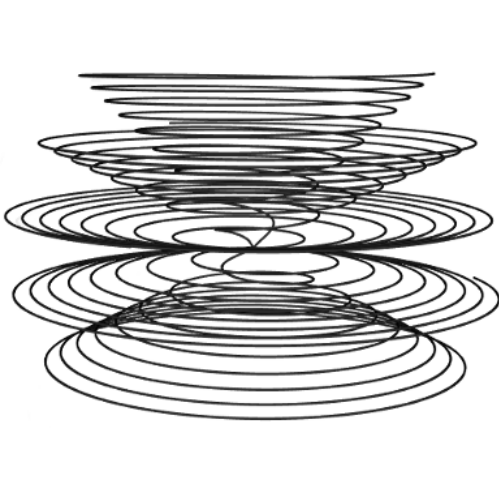}
		\subcaption{\label{subfig:non_cartesian_floret}}
	\end{minipage}
	\hspace{0.1cm}
	\begin{minipage}{0.27\textwidth}
		\includegraphics[width=\textwidth]{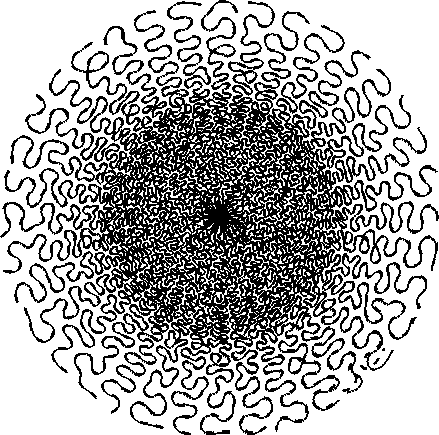}
		\subcaption{\label{subfig:SPARKLING}}
	\end{minipage}
	\caption{\label{fig:non_cartesian_trajectories} Examples of some non-Cartesian trajectories used in CS to accelerate the acquisition:  (a) radial trajectory \cite{Lau}, (b) spiral trajectory, (c) TWIRL: the first combination of radial and spiral trajectories for 2D acquisitions \cite{Jetal}, (d) TWIST: a 3D extension of the TWIRL method \cite{Betal}, (e) FLORET: a 3D non-Cartesian sampling pattern based on the combination of radial and spiral, (f) SPARKLING: a trajectory approximating a target variable density and satisfying physical constraints \cite{L}.}
\end{figure}

Instead of drawing rows of $ A_0 $, which corresponds to probing isolated points in the frequency domain~(k-space) in the context of MRI, it was proposed in~\cite{BBW, ABB} to draw blocks of rows, which corresponds to drawing independent Cartesian lines in k-space. This framework does not cover, however, the case of non-Cartesian acquisition, e.g. acquisition along radial spokes, whose intersection is given by the center of k-space, or more complex trajectories often used in MRI (spiral or non-parametric SPARKLING trajectories, see Figure \ref{fig:non_cartesian_trajectories}).

One important aspect of the non-Cartesian setting from the CS theory viewpoint is that when frequencies of the Fourier transform are not taken to be in $ \mathbb{Z}^d $, then the corresponding matrix $ A_0 $ no longer fulfills the condition $ A_0^* A_0 = I $. The isotropy condition is violated in the non-Cartesian setting leading to the necessity to develop a theory for \emph{anisotropic} CS. 

Some classical CS results were extended to the anisotropic setting in  \cite{KG}. The authors provided a theoretical bound on the number of measurements necessary for the exact reconstruction of a sparse vector $ x $ in the case of uniform isolated measurements. Another recent paper \cite{AS} on anisotropic CS extends the results of \cite{KG} to the infinite-dimensional setting.

In the present work we propose to combine the approaches of \cite{KG} and \cite{BBW, ABB} to develop anisotropic CS results that take structured sparsity and variable density structured acquisition into account.

The present note is organised as follows. In section 2 we introduce the notation. In section 3 we present the main result. In section 4 we give the proof of the main theorem. In section 5 we present formulas for optimal sampling densities in the case of isolated measurements and block-structured sampling. The Appendix contains some classical results of probability theory and compressed sensing theory that are used in the proofs of section 4. 

\medskip

\noindent \textbf{Acknowledgements.} This work was carried out during a one-year Inria delegation of \mbox{A.~Kazeykina} in the Parietal team of Inria at NeuroSpin, CEA Saclay.

\section{Notation}
\label{notation_section}
Let $ n \in \mathbb{N} $ and $ d_k \in \mathbb{N} $, $ k = 1, \ldots, M $, such that $ \sum_{ k = 1 }^M d_k = n $. Let $ \mathcal{B}_k \in \mathbb{C}^{ d_k \times n } $ and construct matrix $ A_0 \in \mathbb{C}^{ n \times n } $ by stacking the blocks $ \mathcal{B}_k $ on top of each other: $ A_0 = ( \mathcal{B}_k )_{ k = 1 }^M $. Matrix $ A_0 $ represents the set of possible measurements imposed by a specific sensor device. We will assume that $ A_0^* A_0 $ is invertible.
 
The sensing matrix $ A $ is constructed by drawing randomly blocks of rows of matrix $ A_0 $. More precisely, let $ B $ be a random variable taking values $ \mathcal{B}_k / \sqrt{ \pi_k } $ with probabilities $ \pi_k $, $ k = 1, \ldots, M $. For $ m \leq M $ let $ B_1, \ldots, B_m $ be i.i.d. copies of the random block $ B $. The random sensing matrix is constructed as follows:  
\begin{equation}
\label{A_def}
A = \frac{ 1 }{ \sqrt{ m } } ( B_l )_{ l = 1 }^{ m }.
\end{equation}

Define
\begin{equation*}
X = (\mathbb{ E } [ B^* B ])^{-1}.
\end{equation*}
Note that $ X $ exists since $ \mathbb{ E } [ B^* B ] = A_0^* A_0 $ that we assumed to be invertible. 
Note also that if $ A_0 $ satisfies the isotropy condition $ A_0^* A_0 = I_n $, where $ I_n \in \mathbb{C}^{ n \times n } $ is the identity matrix, then $ X = I_n $.

Let $ S \subset \{ 1, 2, \ldots, n \} $. Denote $ S^c = \{ 1, \ldots, n \} \backslash S $. Define $ P_S $ to be the matrix of the linear projection $ x \mapsto P_S x $, where $ P_S x $ is the restriction of $ x $ to the components in $ S $. Define quantities $ \Theta_S $, $ \Lambda_S $ to be positive numbers such that   
\begin{align}
\label{Theta} & \Theta_S \geq \sqrt{ \| B^* B X P_S^* \|_{ \infty \to \infty } \| P_S X B^* B \|_{ \infty \to \infty } } \text{ a.s.} \\
\label{Lambda}& \Lambda_S \geq \| P_S X B^* B P_S^* \|_{ 2 \to 2 }  \text{ a.s.}
\end{align}

Note that
\begin{multline*}
\| P_S X B^* B P^*_S \|_{ 2 \to 2 } \leq \sqrt{ \| P_S X B^* B P^*_S \|_{ 1 \to 1 } \| P_S X B^* B P^*_S \|_{ \infty \to \infty } } = \\
\sqrt{ \| P_S B^* B X P^*_S \|_{ \infty \to \infty } \| P_S X B^* B P^*_S \|_{ \infty \to \infty } } \leq \sqrt{ \| B^* B X P^*_S \|_{ \infty \to \infty } \| P_S X B^* B \|_{ \infty \to \infty } }\leq \Theta_S,
\end{multline*}
and thus, if $ \Lambda_S $ is taken as the least upper-bound, then
\begin{equation}
\label{lambda_theta}
\Lambda_S \leq \Theta_S.
\end{equation}

Note that in the isotropic case ($ X = I_n $) $ \Theta_S $ defined by (\ref{Theta}) does not coincide with the quantity $ \Theta( S ) $ introduced in \cite{BBW, ABB}. That is due to the fact that a straightforward generalisation of the definition used in \cite{BBW, ABB} to the anisotropic case does not preserve the relation (\ref{lambda_theta}) verified in the isotropic case. To preserve this relation in the anisotropic case we prefer to consider a symmetrised version of $ \Theta $.

We will denote by $ (e_i)_{ i = 1 }^{ n } $ the canonical basis of $ \mathbb{R}^n $. 

Finally, for a number $ x \in \mathbb{C} $ we denote
\begin{equation*}
\sgn( x ) = \begin{cases} & \frac{ x }{ | x | }, \; x \neq 0, \\ & 0, \; x = 0 \end{cases}
\end{equation*}
and for a vector $ x \in \mathbb{C}^n $ we define $ \sgn( x ) = ( \sgn( x_j ) )_{ j = 1 }^n $.

\section{Main result}

\begin{theorem}
\label{ex_reconstr_theorem}
Let $ x \in \mathbb{R}^n $ or $ \mathbb{C}^n $ be a vector supported on $ S $, such that $ \sgn( x_S ) $ forms a Rademacher or a Steinhaus sequence. Let $ A $ be the random sensing matrix defined by (\ref{A_def}) associated with parameter $ \Theta_S $. Suppose we are given the data $ y = A x $. Then, given $ 0 < \varepsilon < 1 $ and provided that
\begin{align}
\label{m_bound}
& m > c\, \Theta_S ( \Theta_S + 2 ) \ln^2 \left( \frac{ 8 n }{ \varepsilon } \right)
\end{align} 
for $  c $ a numerical constant, the vector $ x $ is the unique minimizer of the basis pursuit problem (\ref{BPP}) with probability at least $ 1 - \varepsilon $. 
\end{theorem}

\begin{rem}
Note that the analogous result in the isotropic case (see Theorem 3.3 of \cite{ABB}) ensures the exact reconstruction of $ x $ provided that $ m \gtrsim \Theta \ln^2\left( 6 n / \varepsilon \right) $. The possibility to obtain a bound in $ \Theta $ rather than in $ \Theta^2 $ is due to the presence of additional symmetries that can be efficiently exploited in the isotropic case (see also Remark \ref{rem_lemmas}).
\end{rem}

\section{Proof}
The proof of Theorem \ref{ex_reconstr_theorem} is based on the following proposition.
\begin{prop}
\label{prop}
For $ x \in \mathbb{C}^N $ with support $ S $ if 
\begin{itemize}
\item[(i)] $ P_S X A^* A P^*_S $ is injective,
\item[(ii)] $ \left| \langle ( P_S X A^* A P^*_S )^{-1} P_S X A^* A e_l , \sgn( x_S ) \rangle \right| < 1 $ $ \forall l \in S^c $,
\end{itemize}
then the vector $ x $ is the unique solution of (\ref{BPP}). 
\end{prop}
\noindent\emph{Proof}
This proposition is a corollary of Theorem 4.26 of \cite{FR} that we formulate as Theorem \ref{FRtheorem} of the Appendix (see also Corollary 4.28 of \cite{FR}).

Indeed, consider the condition (ii) of Theorem \ref{FRtheorem}. 
First of all, we note that the fact that $ P_S X A^* A P^*_S $ is injective implies that $ A P^*_S $ is injective.
Next, take 
\bee
h = A X P^*_S ( P_S A^* A X P^*_S )^{-1} \sgn( x_S )
\ene
where $ ( P_S A^* A X P^*_S )^{-1} $ exists due to assumption (i) of the proposition (indeed $ P_S A^* A X P^*_S $ is the adjoint of the matrix $ P_S X A^* A P^*_S $ which is invertible because it is square and injective).

Then the condition $ P_S A^* h = \sgn( x_S ) $ is satisfied.
Further, the condition $ | ( A^* h )_l | < 1 $, $ l \in S^c $ is rewritten as $ | \langle A e_l , h \rangle | < 1 $, $ l \in S^c $, which is satisfied if (ii) of Proposition \ref{prop} is satisfied. 

\qed

We now formulate and prove two Lemmas that will be used in the proof of Theorem \ref{ex_reconstr_theorem}.

\begin{lemma}
\label{loc_is_lemma}
For every $ S \subset \{ 1, 2, \ldots, n \} $ with $ |S| = s $ and for every $ \delta > 0 $, the following holds
\begin{equation*}
\mathbb{P}( \| P_S X A^* A P^*_S - I_s \|_{ 2 \to 2 } \geq \delta ) \leq 2s \exp\left( -\frac{ m \delta^2 }{ 4 \Lambda_S ( 2 \Lambda_S + \delta / 3 ) }. \right)
\end{equation*}
\end{lemma}

\noindent\emph{Proof.} Let
\begin{equation*}
M_{i} =  P_S X B_i^* B_i P^*_S \text{ and } X_i := \frac{ 1 }{ m } \left( M_{ i } - \mathbb{E} M_{ i } \right), \quad i = 1, 2, \ldots, m.
\end{equation*}
Then $ P_S X A^* A P^*_S - I_s = \sum_{ i = 1 }^m X_i $.

We have the following estimate:
\begin{equation*}
\| X_i \|^2_{ 2 \to 2 } = \frac{ 1 }{ m^2 }\sup_{ \| x \|_2 \leq 1 } \| ( M_i - \mathbb{E} M_i ) x  \|_2^2 \leq \frac{ 4 \Lambda_S^2 }{ m^2 } =: K^2.
\end{equation*}

Consider $ \sum_{ i = 1 }^{ m } \mathbb{E} X_i^* X_i $, this matrix being self-adjoint we have 
\begin{equation*}
\sigma_1^2 := \| \sum_{ i = 1 }^{ m }  \mathbb{E} X_i^* X_i \|_{ 2 \to 2 } = \sup_{ \| x \|_2 \leq 1 } \sum_{ i = 1 }^{ m } \langle x, \mathbb{E} X_i^* X_i x  \rangle.
\end{equation*}
Since 
\begin{equation*}
\langle x, X_i^* X_i x \rangle \leq \frac{4 \Lambda_S^2}{ m^2 },
\end{equation*}
we have that $ \sigma_1^2 \leq \frac{ 4 \Lambda_S^2 }{ m } $. In a similar way, 
\begin{equation*}
\sigma_2^2 := \| \sum_{ i = 1 }^{ m }  \mathbb{E} X_i X_i^*  \|_{ 2 \to 2 } \leq \frac{ 4 \Lambda_S^2 }{ m }.
\end{equation*}
Finally, the required result follows from Proposition \ref{mB} of Appendix; it suffices to set $ t = \delta $, $ B = K = \frac{ 2 \Lambda_S }{ m } $. 
\qed

\begin{lemma}
\label{aux_un_inc_lemma}
Let $ S \subset \{ 1, 2, \ldots, n \} $. Then, for every $ t > 0 $
\begin{equation*}
\mathbb{P}( \max_{ i \in S^c } \| P_S X A^* A e_i \|_2 \geq \Theta_S / \sqrt{ m } + t  ) \leq n \exp\left( - \frac{ m t^2 / 2 }{ 4 \Theta_S^2 + 4 \Theta_S^2 / \sqrt{ m } + 2 \Theta_S t / 3 } \right).
\end{equation*}
\end{lemma}

\noindent\emph{Proof.}
Fix $ i \in S^c $ and define $ N_j = P_S X B_j^* B_j $, $ Y_j = \frac{ 1 }{ m } ( N_j e_i - \mathbb{E} N_j e_i ) $ and $ Z = \| \sum_{ j = 1 }^{ m } Y_j \|_2 $. Note that $ \mathbb{E} Y_j = 0 $.

First, due to the definition of $ \Theta_S $, we can estimate
\begin{multline}
\label{theta_appears}
\| N_j e_i \|^2 = | \langle N_j^* N_j e_i, e_i \rangle | \leq \| N_j^* N_j e_i \|_2 \leq \| N_j^* N_j e_i \|_1 \leq \| N_j^* N_j \|_{ \infty \to \infty } \leq \\ \| N_j^* \|_{ \infty \to \infty }\| N_j \|_{ \infty \to \infty } \leq \Theta_S^2,
\end{multline}
which is due to the following:
\begin{equation*}
\| M \|_{ \infty \to \infty } = \max_{ i } \| e_i^* M \|_1.
\end{equation*}
Thus the following estimate is true: $ \| Y_j \|_2 \leq 2 \Theta_S / m =: K $. 

The next required estimate is obtained by using the Cauchy-Schwarz inequality and (\ref{theta_appears}):
\begin{multline*}
\sup_{ \| x \|_2 \leq 1 } \sum_{ j = 1 }^m \mathbb{E}| \langle x, Y_j \rangle |^2 = \frac{ 1 }{ m^2 } \sup_{ \| x \|_2 \leq 1 } \sum_{ j = 1 }^m \mathbb{E}| \langle x, N_j e_i - \mathbb{E} N_j e_i \rangle |^2 \leq \\  \leq \frac{ 1 }{ m^2 } \sup_{ \| x \|_2 \leq 1 } \sum_{ j = 1 }^m \mathbb{E} \| x \|_2^2 ( \| N_j e_i \|_2 + \mathbb{E} \| N_j e_i \|_2 )^2  \leq 
\frac{ 4 \Theta_S^2 }{ m } =: \sigma^2.
\end{multline*}

We use the independence of the vectors $ Y_j $ and the fact that they have zero mean value to get the following estimate:
\begin{equation*}
( \mathbb{E} Z )^2 \leq \mathbb{E} Z^2 = \mathbb{E} \| \sum_{ j = 1 }^{ m } Y_j \|_2^2 = \sum_{ j = 1 }^{ m } \mathbb{E} \| Y_j \|_2^2 + \sum_{ j = 1 }^{ m } \sum_{ k \neq j } \langle \mathbb{E} Y_j, \mathbb{E} Y_k \rangle = \sum_{ j = 1 }^{ m } \mathbb{E} \| Y_j \|_2^2.
\end{equation*}
Now we use (\ref{theta_appears}) again to estimate:
\begin{multline*}
\sum_{ j = 1 }^{ m }  \mathbb{E} \| Y_j \|^2_2 =
\frac{ 1 }{ m^2 } \sum_{ j = 1 }^{ m } \mathbb{E} \| N_j e_i - \mathbb{E} N_j e_i \|^2_2 = \frac{ 1 }{ m^2 } \sum_{ j = 1 }^{ m } ( \mathbb{E} \| N_j e_i \|^2_2 - \| \mathbb{E} N_j e_i \|_2^2 ) \leq 
\frac{ 1 }{ m^2 } \sum_{ j = 1 }^{ m } \mathbb{E} \| N_j e_i \|^2_2
\leq \frac{ \Theta_S^2 }{ m },
\end{multline*}
which implies 
\begin{equation*}
\mathbb{E} Z \leq \sqrt{ \Theta_S^2 / m } =: \mu.
\end{equation*}
The result then follows from Proposition \ref{vBv1} of Appendix and a union bound. 

\qed 

\begin{rem}\label{rem_lemmas}

Note that for $ X = I_n $ the estimates of Lemmas \ref{loc_is_lemma}, \ref{aux_un_inc_lemma} are looser than those obtained for the isotropic case in \cite{ABB} (see Lemmas C.1, C.2 of \cite{ABB}; note $ \Lambda^2 $ instead of $ \Lambda $ and $ \Theta^2 $ instead of $ \Theta $ in our results). That is due to the fact that in the isotropic case it is possible to exploit some extra symmetries to obtain tighter estimates.
\end{rem}

\noindent\textit{Proof of Theorem \ref{ex_reconstr_theorem}.} We follow the reasoning proposed in \cite{ABB}.

By Lemma \ref{loc_is_lemma}, condition (i) of Proposition \ref{prop} fails with probability not higher than $ 2 s \exp\left( \frac{ m }{ 4 \Lambda_S ( 2 \Lambda_S + 1/3 ) } \right)  $. The latter expression is bounded by $ \frac{ \varepsilon }{ 4 } $ provided that
\bee
m \geq 4 \Lambda_S ( 2 \Lambda_S + \frac{1}{3} ) \ln \left( \frac{ 8 s }{ \varepsilon } \right).
\ene

Now let us study when condition (ii) of Proposition \ref{prop} fails. Denote $ A_S^{\dagger} = ( P_S X A^* A P^*_S )^{-1} P_S X A^* $. Then, by union bound,
\begin{multline*}
\begin{aligned}
& \mathbb{P}( \text{(ii) fails } ) = \mathbb{P}\left( \exists \, l \in S^c \colon  \left| \langle A_S^{\dagger} A e_l , \sgn( x_S ) \rangle \right|  \geq 1 \right) \\
& \leq \mathbb{P}\left( \exists l \, \in S^c \colon  \left| \langle A_S^{\dagger} A e_l , \sgn( x_S ) \rangle \right|  \geq 1 \text{ and } \max_{ l \in S^c } \| A_S^{ \dagger } A e_l \|_2 \leq \alpha \right) + \mathbb{P}\left( \max_{ l \in S^c } \| A_S^{ \dagger } A e_l \|_2 \geq \alpha \right) \\
& \leq \sum_{ l \in S^c }  \mathbb{P}\left( \left| \langle A_S^{\dagger} A e_l , \sgn( x_S ) \rangle \right|  \geq \alpha^{-1} \| A_S^{ \dagger } A e_l \|_2  \text{ and } \max_{ l \in S^c } \| A_S^{ \dagger } A e_l \|_2 \leq \alpha \right) + \mathbb{P}\left( \max_{ l \in S^c } \| A_S^{ \dagger } A e_l \|_2 \geq \alpha \right) \\
& \leq 2 n \exp\left( - \frac{ 1 }{ 2 \alpha^2 } \right) + \mathbb{P}\left( \max_{ l \in S^c } \| A_S^{ \dagger } A e_l \|_2 \geq \alpha \right),
\end{aligned}
\end{multline*}
where the last bound is due to Hoeffding type inequality for Rademacher or Steinhaus sequence (see Propositions \ref{hoeff_rademacher}, \ref{hoeff_steinhaus}: we take $ a_l $ to be equal to $ ( P_S X A^* A P^*_S )^{-1} P_S X A^* A e_l $ and we set $ u = \alpha^{-1} $, $ \lambda = \frac{1}{2} $ ).

Now we study the second term. Note that
\bee
\| A_S^{ \dagger } A e_l \|_2 = \| ( P_S X A^* A P^*_S )^{-1} P_S X A^* A e_l \|_2 \leq \| ( P_S X A^* A P^*_S )^{-1} \|_{2 \to 2} \| P_S X A^* A e_l \|_2.
\ene

Take $ 0 < \delta < 1 $ and $ \tilde t > 0 $. Denote $ s = | S | $. Let $ \mathcal{A} $ be the event that $ \| P_S X A^* A P^*_S - I_s \|_{ 2 \to 2 } < \delta  $ and let $ \mathcal{B} $ be the event that $ \max_{ i \in S^c } \| P_S X A^* A e_i \|_2 < \tilde{t} $. Note that $ \mathcal{A} $ implies that $ \| ( P_S X A^* A P^*_S )^{-1} \| < \frac{ 1 }{ 1 - \delta }  $. 

Set $ \alpha = \frac{\tilde t}{ 1 - \delta } $.Then $ \mathcal{A} \cap \mathcal{B} $ implies that $ \max_{ l \in S^c } \| A_S^{ \dagger } A e_l \|_2 < \alpha $, and so $ \max_{ l \in S^c } \| A_S^{ \dagger } A e_l \|_2 \geq \alpha $ means that $ \mathcal{A}^c \cup \mathcal{B}^c $ holds, where $ \mathcal{E}^c $ denotes the event complementary to $ \mathcal{E} $. Thus we get the following estimate
\bee
\mathbb{P}( \max_{ l \in S^c } \| A_S^{ \dagger } A e_l \|_2 \geq \alpha ) \leq \mathbb{P}( \| P_S X A^* A P^*_S - I_s \|_{ 2 \to 2 } \geq \delta ) + \mathbb{P}( \max_{ i \in S^c } \| P_S X A^* A e_i \|_2 \geq \tilde{t} ).
\ene
Define 
\bee
P_1 = 2 n \exp\left( - \frac{ 1 }{ 2 \alpha^2 } \right), \quad P_2 = \mathbb{P}( \| P_S X A^* A P^*_S - I_s \|_{ 2 \to 2 } \geq \delta ), \quad P_3 = \mathbb{P}( \max_{ i \in S^c } \| P_S X A^* A e_i \|_2 \geq \tilde{t} ).
\ene

By Lemma \ref{loc_is_lemma} the probability $ P_2 $ is bounded by $ 2s \exp\left( - \frac{ m \delta^2 }{ 4 \Lambda_S ( 2 \Lambda_S + \delta/3 ) } \right) $. Thus it can be majorised by $ \frac{\varepsilon}{4} $ if 
\begin{equation}
\label{m_lambda_estimate}
m \geq \frac{ 4 \Lambda_S ( 2 \Lambda_S + \delta / 3 ) }{ \delta^2 } \ln \left( \frac{ 8 s }{ \varepsilon } \right).
\end{equation}

Now take $ \tilde t = \frac{\Theta_S}{\sqrt{m}} + t $ for some $ t > 0 $. By Lemma \ref{aux_un_inc_lemma} probability $ P_3 $ is bounded by $ \frac{\varepsilon}{4} $, if 
\bee
m \geq \frac{8}{t^2} \Theta_S ( \Theta_S + \Theta_S / \sqrt{m} + t / 6 ) \ln \left( \frac{ 4 n }{ \varepsilon } \right).
\ene
If we assume that $ m \geq \Theta_S^2 $, then we can write that $ P_3 $ is bounded by $ \frac{\varepsilon}{4} $, if 
\begin{equation}
\label{m_theta_estimate}
m \geq \frac{8}{t^2} \Theta_S ( \Theta_S + 1 + t / 6 ) \ln \left( \frac{ 4 n }{ \varepsilon } \right).
\end{equation}

Finally, set $ t = \delta $, assume 
\begin{equation}
\label{m_assumption}
m \geq c \Theta_S^2 \ln \left( \frac{ 8n }{ \varepsilon } \right), \text{ for some constant } c > 0,
\end{equation}
and choose $ \delta = \sqrt{ \frac{ 1 }{ c' \ln \left( \frac{ 8 n }{ \varepsilon } \right) } } $ with $ \min( c', c ) \geq 16 $. Then $ P_1 $ is bounded by $ \frac{ \varepsilon }{ 4 } $ if 
\bee
2 n \exp \left( - \frac{ ( 1 - \delta )^2 }{ 2 ( \Theta_S / \sqrt{m} + \delta )^2 } \right) \leq \frac{ \varepsilon }{ 4 } \quad \Leftrightarrow \quad \frac{ ( 1 - \delta )^2 }{ 2 ( \Theta_S / \sqrt{m} + \delta )^2 } \geq \ln \left( \frac{ 8 n }{ \varepsilon } \right).
\ene 
The latter condition is satisfied due to assumption (\ref{m_assumption}) and the choice of $ \delta $. Indeed, due to (\ref{m_assumption}) we have that 
\bee
\frac{ \Theta_S }{ \sqrt{m} } \leq \frac{ 1 }{ \sqrt{ c \ln \left( \frac{ 8 n }{ \varepsilon } \right) } }
\ene 
and thus, due to the definition of $ \delta $,
\bee
\left( \frac{ \Theta_S }{ \sqrt{m} } + \delta \right)^2 \leq \frac{ 4 }{ \min( c, c' ) \ln \left( \frac{ 8n }{ \varepsilon } \right) },
\ene
which implies that 
\bee
\frac{ ( 1 - \delta )^2 }{ 2 ( \Theta_S / \sqrt{m} + \delta  )^2 } \geq \frac{ \min( c, c' ) }{ 8 } \ln \left( \frac{ 8 n }{ \varepsilon } \right) ( 1 - \delta )^2 \geq \ln \left( \frac{ 8 n }{ \varepsilon } \right).
\ene

Plugging the chosen value of $ \delta $ into (\ref{m_lambda_estimate}), we obtain that it suffices to take 
\begin{align*}
& m \geq 64 \Lambda_S ( 2 \Lambda_S + 1 )  \ln \left( \frac{ 8n }{ \varepsilon } \right) \ln \left( \frac{ 8s }{ \varepsilon } \right), \\
& m \geq 128 \Theta_S ( \Theta_S + 2 ) \ln^2 \left( \frac{ 8n }{ \varepsilon } \right)
\end{align*}
for (BP) to have a unique solution with probability $ 1 -\varepsilon $. If we choose $ \Lambda_S $ in the definition (\ref{Lambda}) to be the least upper-bound, then the inequality (\ref{lambda_theta}) implies that it suffices to choose $ m $ verifying (\ref{m_bound}) to guarantee the result of Theorem \ref{ex_reconstr_theorem}.

\qed

\section{Optimal sampling strategies}
In this Section we derive formulas for probabilities $ \pi_k $ minimising the quantity $ \Theta_S $ defined by (\ref{Theta}) and arising in the bound (\ref{m_bound}). 

We consider the following two principal sampling strategies.
\begin{itemize}
\item \textbf{Isolated measurements} 

Let $ ( a_i^* )_{ 1 \leq i \leq n } \in \mathbb{C}^n $ be a set of row vectors. Set $ M = n $, $ d_k = 1 $ for all $ k $ and $ \mathcal{B}_k = a_k^* $. This setting represents isolated measurements in MRI.

\item \textbf{Block-structured sampling}
Let $ ( a_i^* )_{ 1 \leq i \leq n } \in \mathbb{C}^n $ be a set of row vectors and let $ ( \mathcal{I}_k )_{ 1 \leq k \leq M } $ denote a partition of the set $ \{ 1, \ldots, n \} $. The rows $ ( a_i^* ) $ are then partitioned into blocks $ ( D_k )_{ 1 \leq k \leq M } $: $ D_k = ( a^*_i )_{ i \in \mathcal{I}_k } $. In this setting $ d_k = | \mathcal{I}_k | $ and $ \mathcal{B}_k = D_k $. This setting corresponds to sampling blocks of measurements in MRI. 

\end{itemize}

Define 
\bee
c^1_{S,k} = \| a_k \|_{ \infty } \| a_k^* X P^*_S \|_1, \quad c_{S,k}^2 = \| P_S X a_k \|_{ \infty } \| a_k^* \|_1, \quad k = 1, \ldots, n.
\ene

\begin{prop}
\label{prop_isol}
In the setting of isolated measurements the probability minimising $ \Theta_S $ is
\bee
\pi^{ \Theta_S }_k = \frac{ \sqrt{ c^1_{ S, k }  c^2_{ S, k } }  }{ \sum_{k = 1}^{ n }  \sqrt{ c^1_{ S, k }  c^2_{ S, k } }  }, \quad k = 1, 2, \ldots, n.
\ene
The corresponding $ \Theta_S $ is given by
\bee
\Theta_S = \sum_{k = 1}^{ n }  \sqrt{ c^1_{ S, k }  c^2_{ S, k } }.
\ene
\end{prop}

Define
\bee
C^1_{ S, k }  = \| D_k^* D_k X P^*_S \|_{ \infty \to \infty }, \quad 
C^2_{ S, k }  =  \| P_S X D_k^* D_k \|_{ \infty \to \infty }, \quad k = 1, \ldots, M.
\ene

\medskip

\begin{prop}
\label{prop_block}
In the setting of block-structured sampling the probability minimising $ \Theta_S $ is
\bee
\pi^{ \Theta_S }_k = \frac{ \sqrt{ C^1_{ S, k }  C^2_{ S, k } }  }{ \sum_{k = 1}^{ M }  \sqrt{ C^1_{ S, k }  C^2_{ S, k } } }, \quad k = 1, 2, \ldots, M.
\ene
The corresponding $ \Theta_S $ is given by
\bee
\Theta_S = \sum_{k = 1}^{ M }  \sqrt{ C^1_{ S, k }  C^2_{ S, k } }.
\ene
\end{prop}

The proof of Propositions \ref{prop_isol} and \ref{prop_block} follows from Lemma D1 of \cite{ABB}.

\begin{rem}
Note that for $ X = I_n $ we do not find the same expressions for $ \pi^{ \Theta_S } $ as those presented in \cite{ABB} for the isotropic case, but rather their symmetrised versions. That difference is due to the difference in the definition of $ \Theta $ explained in Section \ref{notation_section}.
\end{rem}

\section{Conclusion}
In the anisotropic setting we provided a theoretical bound on the number of measurements that are needed to reconstruct a sparse signal with large probability. The bound is given in terms of a quantity that allows to take into account a priori information on the sparsity structure of the signal and to apply variable density block-structured sampling strategies.
We have provided general formulas for probability distributions optimising the obtained theoretical bound. We hope that these formulas can be further analysed to derive explicit expressions for optimal sampling densities in the context of non-Cartesian MRI.

\section*{Appendix}

\begin{prop}(Matrix Bernstein inequality \cite{T}) 

\label{mB}
Consider a finite sequence $ \{ M_k \} \in \mathbb{C}^{ d \times d } $ of independent random matrices. Assume that each random matrix satisfies $ \mathbb{E}[ M_k ] = 0 $ and $ \| M_k \| \leq B $ a.s. and define 
\begin{equation*}
\sigma^2 = \max \left\{ \| \sum_{k} \mathbb{E}( M_k M_k^* ) \|_{2 \to 2}, \| \sum_k \mathbb{E} ( M_k^* M_k ) \|_{ 2 \to 2 } \right\}.
\end{equation*}
Then for all $ t \geq 0 $,
\begin{equation*}
\mathbb{P} \left( \| \sum_k M_k \|_{ 2 \to 2 } \geq t \right) \leq 2d \exp\left( - \frac{ t^2 / 2 }{ \sigma^2 + Bt / 3 } \right).
\end{equation*}

\end{prop}

\begin{prop}(Vector Bernstein inequality \cite{ABB}) 

\label{vBv1}
Consider a set of independent random vectors $ Y_1, Y_2, \ldots, Y_m $ such that 
\begin{equation*}
\mathbb{E} Y_i = 0, \quad \| Y_i \|_2 \leq K \text{ a.s. } \forall i = 1, \ldots, m,
\end{equation*}
and let $ \sigma, \mu > 0 $ such that 
\begin{equation*}
\sup_{ \| x \|_2 \leq 1 } \sum_{ i = 1 }^{ m } \mathbb{E} | \langle x, Y_i \rangle  |^2 \leq \sigma^2, \quad \mathbb{E} Z \leq \mu, \quad \text{ where  } Z : = \| \sum_{ i = 1 }^{ m } Y_i \|_2.
\end{equation*} 
Then for every $ t > 0 $ the following holds:
\begin{equation*}
\mathbb{P}( Z > \mu + t ) \leq \exp\left( - \frac{ t^2 / 2 }{ \sigma^2 + 2 K \mu + t K / 3 } \right)
\end{equation*}
\end{prop}

\begin{prop}(Hoeffding's bound for Rademacher sequence \cite{FR})
\label{hoeff_rademacher}
Let $ a \in \mathbb{C}^M $ and $ \epsilon = ( \epsilon_1, \epsilon_2, \ldots, \epsilon_M ) $ be a Rademacher sequence.Then
\begin{equation*}
\mathbb{P}\left( \sum_{ i = 1 }^{ M } | \epsilon_i a_i | \geq u \| a \|_2 \right) \leq 2 \exp( -u^2 / 2 ) \quad \forall u > 0.
\end{equation*}
\end{prop}

\begin{prop}(Hoeffding-type bound for Steinhaus sequence \cite{FR})
\label{hoeff_steinhaus}
Let $ a \in \mathbb{C}^M $ and $ \epsilon = ( \epsilon_1, \epsilon_2, \ldots, \epsilon_M ) $ be a Steinhaus sequence.Then for any $ 0 < \lambda < 1 $
\begin{equation*}
\mathbb{P}\left( \sum_{ i = 1 }^{ M } | \epsilon_i a_i | \geq u \| a \|_2 \right) \leq \frac{ 1 }{ 1 - \lambda } \exp( - \lambda u^2 ) \quad \forall u > 0.
\end{equation*}
\end{prop}

\begin{theorem}{(Theorem 4.26 of \cite{FR})}
\label{FRtheorem}
Given a matrix $ A \in \mathbb{C}^{m \times N} $, a vector $ x \in \mathbb{C}^N $ with support $ S $ is
the unique minimizer of $ \| z \|_1 $ subject to $ Az = Ax $ if one of the following equivalent conditions holds:
\begin{itemize}
\item[(i)]
\bee
\left| \sum\limits_{ j \in S }
\overline{\sgn(x_j)} v_j \right| < \| v_{S^c} \|_1 \text{ for all } v \in \ker A \backslash \{0\},
\ene
\item[(ii)] $ A P^*_S $ is injective and there exists a vector $ h \in \mathbb{C}^m $ such that
\bee
(A^*h)_j = \sgn(x_j), \; j \in S, \quad |(A^*h)_l|< 1, \; l \in S^c.
\ene
\end{itemize}
\end{theorem}

\end{document}